\documentclass[a4paper]{jpconf}
\pdfoutput=1

\usepackage{iopams}
\usepackage{amssymb}
\usepackage{amsmath}

\usepackage{url}

\usepackage{graphicx}
\usepackage{hyperref}
\usepackage{color}

\def\commitID{commitID: a5baf4b78648652b8add5cff0deb0875a6de62b7}
\def\commitDATE{ Fri Mar 2 16:52:05 2012 +0100}

\def\commitSTATUS{UNCLEAN}

\newcommand{\dcc}{LIGO-P1100172-v3}

\newcommand{\ltitle}{Coherent follow-up of Continuous Gravitational Wave
candidates: minimal required observation time}

 \hypersetup{
   pdftitle={\ltitle},
   pdfauthor={Miroslav Shaltev},
   pdfkeywords={\$\commitID-\commitSTATUS{} \$}
 }

\newcommand{\mF}{\mathcal{F}}
\newcommand{\mFth}{\mathcal{F}_{\mathrm{th}}}
\newcommand{\mV}{\mathcal{V}}
\newcommand{\Tcoh}{T}

\newcommand{\Tsft}{T_{\mathrm{SFT}}}

\newcommand{\Nseg}{N}
\newcommand{\Nsft}{N_{\mathrm{SFT}}}
\newcommand{\Tseg}{\Delta T}
\newcommand{\Ntemp}{\mathcal{N}}
\newcommand{\Ctot}{C}

\newcommand{\Nsky}{\Ntemp_{sky}}

\newcommand{\snr}{\rho}
\newcommand{\snrcand}{\snr_{\mathrm{c}}}
\newcommand{\snrth}{\snr_{\mathrm{th}}}
\newcommand{\snrac}{\snr_{\mathrm{ac}}}
\newcommand{\Nd}{N_{\mathrm{d}}}
\newcommand{\Ndcand}{\Nd^{\mathrm{c}}}

\newcommand{\thickness}{\theta}

\newcommand{\osb}{V_{0}}

\newcommand{\sfp}{h}

\newcommand{\colevel}{\kappa}

\newcommand{\mFn}{\mF_{\Ntemp}}

\newcommand{\mFc}{\mF_{\mathrm{c}}}

\newcommand{\Hz}{\mathrm{Hz}}
\newcommand{\second}{\mathrm{s}}
\newcommand{\years}{\mathrm{yr}}
\newcommand{\radian}{\mathrm{rad}}

\newcommand{\detect}{\mathrm{det}}
\newcommand{\FA}{\mathrm{fA}}
\newcommand{\FD}{\mathrm{fD}}

\newcommand{\sep}{\mathrm{sep}}

\newcommand{\pdet}{p_{\detect}}
\newcommand{\pjoint}{p_{J}}

\newcommand{\psep}{p_{\sep}}
\newcommand{\pFA}{p_{\FA}}
\newcommand{\pFD}{p_{\FD}}

\begin{document}

\title{Coherent follow-up of Continuous Gravitational-Wave
candidates: minimal required observation time}

\author{Miroslav Shaltev}


\ead{miroslav.shaltev@aei.mpg.de}
\address{Albert-Einstein-Institut, Callinstr.\ 38, 30167 Hannover, Germany}

\address{\dcc \qquad\commitDATE}

\begin{abstract}
We derive two different methods to compute the minimal required integration time
of a fully coherent follow-up of candidates produced in wide parameter space 
semi-coherent searches, such as global correlation StackSlide searches using
Einstein@Home. We numerically compare these methods in terms of integration
duration and computing cost. In a Monte Carlo study we confirm that we can
achieve the required detection probability.

\end{abstract}

\section{Introduction}
\label{sec:1}
Isolated neutron stars as potential sources of continuous gravitational waves
are optimally studied with fully coherent matched filtering  methods. These
methods are not directly applicable to previously unknown objects due to the
large parameter space that needs to be covered in all-sky wide parameter space 
searches and the related enormous computing cost \cite{Brady:1997ji}. Advanced
semi-coherent techniques, e.g. StackSlide searches on the distributed computing
environment Einstein@Home \cite{EAH:Misc}, produce candidates that require
follow-up in greatly reduced parameter space regions. A follow-up scheme
consists of two basic stages. In the first \textit{refinement} stage, we find
the maximum-likelihood estimator and associated optimal search volume $\osb$. In
the second \textit{zoom} stage, we zoom in on the optimal search volume by
semi-coherent or fully-coherent integration. In this paper we focus on a
fully-coherent zoom for which we derive and discuss two different methods to
compute the minimal required coherent integration time in order to distinguish
real signals from noise.

\section{Properties of $\mF$-statistic searches}
\label{sec:2}
The $\mF$-statistic was first derived in \cite{Jaranowski:1998qm} for the single
detector case and generalized to multi-detector searches in
\cite{cutler05:_gen_fstat}. Continuous gravitational-wave signals are
monochromatic and sinusoidal in the frame of the gravitational-wave source and
undergo phase- and amplitude modulation due to the rotation and orbital motion
of the detector. The $\mF$-statistic is analytically amplitude-maximized, thus
the parameter space to search for signals is spanned by the remaining ``Doppler
parameters'' $\lambda$, namely sky position ($\alpha$ - right ascension,
$\delta$ - declination) and intrinsic frequency and frequency derivatives
($f,\dot{f},\ddot{f}$...), further referred to as spindowns. Searching for
previously unknown objects with matched filtering implies computing matched
filters for different points in parameter space, also referred to as templates.
As realized in \cite{bala96:_gravit_binaries_metric,owen96:_search_templates} in
the context of searches for gravitational waves from inspiraling binaries, a
geometrical approach is best suited for optimal template placement and template
counting. This is made possible by the introduction of a metric tensor $g_{ij}$
on the parameter space and mismatch $m$
\begin{equation}
 \label{eq:1}
 m = g_{ij}\Delta\lambda^{i}\Delta\lambda^{j} + \mathcal{O}(\Delta\lambda^{3})\
,
\end{equation}
where the mismatch $m$ measures the fractional loss of (squared) signal to noise
ratio (SNR) $\snr^{2}$ due to the usage of a nearby template $\lambda_{c}$  with
offset $\Delta\lambda = \lambda_{c} - \lambda_{s}$ from the true parameters of
a putative signal
$\lambda_{s}$
\begin{equation}
 \label{eq:2}
 m = \frac{\snr^{2}_{s}-\snr^{2}_{c}}{\snr^{2}_{s}}\ ,
\end{equation}
with the squared SNR $\snr^{2}_{s}$ and $\snr^{2}_{c}$ obtained at point
$\lambda_{s}$ and $\lambda_{c}$, respectively. Given the metric, the
problem of efficient lattice and alternative random and stochastic template-bank
construction is studied in
\cite{2007arXiv0707.0428P,2009PhRvD..79j4017M,0908.2090v1}.

\subsection{Fully-coherent search}

A fully-coherent search is the classical and most sensitive
$\mF$-statistic-based search in the case of unlimited available computing power
or a sufficiently cheap computing cost requirement. The squared SNR $\snr^{2}$
scales linearly with the observation time $\Tcoh$, according to the following
formula:
\begin{equation}
\label{eq:3}
 \snr^{2} = h_{0}^{2}R\Nd\Tcoh S^{-1}(f)\ ,
\end{equation}
where $h_{0}$ is the intrinsic signal amplitude, $R$ represents the geometrical
``detector response'' , $S$ is the one-sided noise spectral density, which is
assumed constant in a narrow frequency band around $f$, and $\Nd$ is the number
of detectors \cite{prix:_cfsv2}. In the presence of a signal, the
$\mF$-statistic follows a non-central $\chi^2$ distribution with four degrees of
freedom and non-centrality parameter $\snr^{2}$. Thus the expectation value is
\begin{equation}
 \label{eq:4}
E[2\mF_{_{S}}] = 4 + \snr^{2}\ ,
\end{equation}
with standard deviation
\begin{equation}
  \label{eq:5}
\sigma(2\mF_{_{S}}) = \sqrt{2(4+2\snr^{2})}\ .
\end{equation}

\subsection{Semi-coherent search}

At fixed and limited computing cost a more sensitive detection statistic can be
constructed from the incoherent combination of results obtained by coherent
integration of shorter data segments. In particular we consider a Stack-Slide
search \cite{Brady:1998nj,PhysRevD.72.042004,PrixShaltev2011}, where the
statistic is the sum of the $\mF$-statistic over the segments:
\begin{equation}
 \label{eq:6}
 \Sigma = \sum_{k=1}^{\Nseg}2\mF_{k}(\lambda).
\end{equation}
This new statistic $\Sigma$ follows a non-central $\chi^{2}$ distribution with
$4\Nseg$ degrees of freedom, thus the expectation value is
\begin{equation}
 \label{eq:7}
 E[\Sigma] = 4\Nseg + \snr^{2}_{\Sigma}\ ,
\end{equation}
where the non-centrality parameter is the sum of the squared SNRs over different
segments
\begin{equation}
 \label{eq:8}
 \snr^{2}_{\Sigma} = \sum_{k=1}^{\Nseg}\snr^{2}_{k}\ .
\end{equation}
A trivial but useful reformulation of Eq. (\ref{eq:7}) is in terms of average
$2\bar{\mF}=\frac{1}{N}\sum_{k}2\mF_{k}$ and
$\bar{\snr^{2}}=\frac{1}{N}\sum_{k}\snr^{2}_{k}$, namely
\begin{equation}
 \label{eq:9}
 E[2\bar{\mF}] = 4 + \bar{\snr^{2}}\ .
\end{equation}

\subsection{Template counting}

The number of templates sufficient to cover the search volume $\osb$ is given
by \cite{2007arXiv0707.0428P}
\begin{equation}
 \label{eq:10}
 \Ntemp_{n} = \thickness m^{-n/2}\mV_{n}\ ,
\end{equation}
where $\thickness$ is the normalized thickness characterizing the geometric
structure of covering, $m$ is the maximum allowed mismatch , $n$ the number of
dimensions and 
\begin{equation}
 \label{eq:11}
 \mV_{n} = \int d^{n}\lambda\ \sqrt{\det g}\ ,
\end{equation}
is the metric template-bank volume with $g_{ij}$ the parameter space metric.
This is the general form of the template counting formula, which is valid for
arbitrary lattices and curved parameter spaces. In practice, using the flat
metric approximation, where the metric coefficients are constant, we can take
the determinant out of the integral. Moreover, if the parameter space is a
$n$-dimensional ``box'', we can replace the integral over infinitesimal
displacement $d\lambda$ by a product of $n$ ``search bands'' $\Delta\lambda$,
namely
\begin{equation}
 \label{eq:13}
 \mV_{n} = \sqrt{\det g}\prod_{i=1}^{n}\Delta\lambda_{i}\ .
\end{equation}
Follow-up of candidates from semi-coherent searches involves a semi-coherent
metric, shown in \cite{Brady:1998nj,Pletsch:2010a} to be the average of the
metric computed for every segment. The semi-coherent metric allows us to
estimate the search band $\Delta\lambda_{i}$ around the follow-up candidate
using the diagonal elements of the inverse Fisher matrix
\cite{prix06:_searc,PhysRevD.77.042001}, i.e. 
\begin{equation}
 \label{eq:15}
 \Delta\lambda_{i} \equiv \colevel\sqrt{\bar{\Gamma}^{ii}}\ ,
\end{equation}
with
\begin{equation}
 \label{eq:16}
 \bar{\Gamma}^{ii} = \bar{g}^{ii}/\snr^{2}\ ,
\end{equation}
where $\colevel$ defines the confidence level and $g^{ij}$ is the inverse
matrix to $g_{ij}$. In the present work we use an analytical semi-coherent
metric first derived by Pletsch \cite{Pletsch:2010a}. For coherent integration
time longer than a day, but much shorter than a year, the number of sky
templates at fixed frequency $f$ converges to \begin{equation}
 \label{eq:17}
\Nsky = \frac{2\pi^{3}\tau_{\mathrm{E}}^{2}f^{2}}{m}\ ,
\end{equation}
where $\tau_{\mathrm{E}}\approx21\times10^{-3}s$ is the light travel time from
the Earth's center to the detector \cite{Pletsch:2010a}. The semi-coherent
parameter space is finer than the coherent one by a refinement factor $\gamma$.
Using the notion of \textit{refinement per direction} $\gamma_{n}$ we can also
obtain the search bands from the extents of the fully coherent metric, namely
\begin{equation}
 \label{eq:18}
 \Delta\lambda_{i} = \colevel\sqrt{\frac{g^{ii}}{\gamma_{i}^{2}\snr^{2}}}\ .
\end{equation}
For uniformly distributed segments of data without gaps, based on
\cite{Pletsch:2010a} the refinement  factors can be obtained as
\begin{eqnarray}
\label{eq:19-22}
\gamma_{f}&=&1\ ,\\
\gamma_{\dot{f}}&=&\sqrt{5N^2-4}\ ,\\
\gamma_{\ddot{f}}&=&\sqrt{(35N^4-140N^2+108)/3}\ ,\\
\gamma_{\dddot{f}}&=&\sqrt{(105N^8-1260N^6+5012N^4-6160N^2+2304)/(5N^2-4)}\ .
\end{eqnarray}
Finally, for simplicity of the template-bank construction, we use a hyper-cubic
lattice to place templates, though hyper-cubic lattices are in general
suboptimal, compared to better solutions, e.g. $A^{*}_{n}$ lattice. The
normalized thickness for an $n$-dimensional hyper-cubic grid is
\cite{2007arXiv0707.0428P}
\begin{equation}
 \label{eq:23}
\thickness_{n} = n^{n/2}\,2^{-n}\ .
\end{equation}
The proper choice of the number of dimensions that maximizes the number of
templates \cite{Brady:1997ji,Brady:1998nj,PhysRevD.72.042004,PrixShaltev2011}
$\Ntemp$ is:
\begin{equation}
 \label{eq:24}
 \Ntemp = \max_{n}\Ntemp_{n}\ .
\end{equation}

\subsection{Computing cost}

In the follow-up of real candidates, especially weak signal candidates, along
with the constraint of the total amount of available data, the computing cost
constraint
may limit significantly the feasibility of the search. Thus the computing-cost
requirement is of particular interest. There are currently two different
strategies to implement an $\mF$-statistic search code in LIGO's reference
software suite \texttt{lalsuite}\cite{LALSuite:Misc}, namely the SFT-method
based on short Fourier transforms of the data with duration $\Tsft$
\cite{prix:_cfsv2} and the FFT-method based on barycentric resampling
\cite{2010PhRvD..81h4032P}.
Regarding the computational cost, the FFT method is preferable, as the
computational requirement to calculate the $\mF$-statistic, for a single
point in the parameter space, scales only with $\log \Tcoh$, while the cost of
the SFT algorithm scales with $T$. However, for historical reasons the SFT
method is currently still more often used by LIGO/LSC
\cite{Abbott:2007tda,Abbott:2009nc,2010ApJ...722.1504A}, is well tested and we
can use recent timing information. The computing cost of a SFT-based
$\mF$-statistic
search is \begin{equation}
 \label{eq:25}
 \Ctot = \Ntemp c_{0}\Nsft\ ,
\end{equation}
where $\Nsft$ is the number of used SFTs, namely
\begin{equation}
 \label{eq:26}
\Nsft = \Nd\Tcoh/\Tsft
\end{equation}
and $c_{0}$ is the fundamental implementation- and hardware-specific computing
constant per SFT and template.

\section{Minimal required observation time}
\label{sec:3}
The main scope of the present work is to find the minimal required observation
time that guarantees a certain detection probability of a putative signal buried
deep in the detector noise at a certain confidence level by using the
fully-coherent $\mF$-statistic search technique. We consider two different
methods to compute the required integration duration. In method 1, which is
closely related to hypothesis testing, we use the concept of false-alarm and
false-dismissal probability to achieve certain detection probability. This is
the natural way to compute the required integration time. In method 2 we
alternatively use the more intuitive notion of expectation value to find the
observation duration that guarantees the required detection probability.

\subsection{Method 1}
In absence of a signal, the probability density function of the
$\mF$-statistic reduces to a central $\chi^2$-distribution, and the false-alarm
probability is given by
\begin{equation}
 \label{eq:27}
\pFA^{1} = \int_{2\mFth}^{\infty}d(2\mF)\chi^{2}_{4}(2\mF;0)\ ,
\end{equation}
where $\pFA^{1}$ denotes single trial false-alarm probability and
$\chi^{2}_{4}(2\mF;0)$ is the central $\chi^2$-distribution with $4$ degrees of
freedom. The integration of  $\chi^{2}_{4}(2\mF,0)=\frac{1}{2}\mF e^{-\mF}$
yields 
\begin{equation}
\label{eq:29}
 \pFA^{1} = (1+\mFth)e^{-\mFth}\ .
\end{equation}
The overall false-alarm probability of crossing the threshold $2\mFth$ in
$\Ntemp$ trials is
\begin{equation}
 \label{eq:30}
 \pFA = 1 - (1-\pFA^{1})^{\Ntemp}\approx\pFA^{1}\Ntemp ,
\end{equation}
when $\pFA^{1}\Ntemp\ll1$ \cite{Jaranowski:1998qm,Watts:2008qw}, thus
\begin{equation}
 \label{eq:31}
\pFA^{1} = \pFA / \Ntemp\ .
\end{equation}
We cannot solve Eq. (\ref{eq:29}) analytically, but numerical
solution gives a threshold $2\mFth$ value. This allows us to numerically
integrate the false-dismissal probability
\begin{equation}
\label{eq:32}
 \pFD(2\mFth,\rho^2) = \int_{-\infty}^{2\mFth}(d2\mF)\chi^{2}_{4}(2\mF,\rho^2)\
,
\end{equation}
where $\pFD(2\mFth) = 1 - \pdet$, with the desired detection probability
$\pdet$ and $\chi^{2}_{4}(2\mF,\rho^2)$ is the non-central
$\chi^{2}$-distribution with $4$ degrees of freedom and non-centrality
parameter $\rho^2$. At fixed $\pFA^{*}$ and $\pFD^{*}$, using the above
equation, we can compute a threshold SNR
$\rho_{\mathrm{th}}(\pFA^{*},\pFD^{*})$. The required $\Tcoh$ is such that the
inequality
\begin{equation}
\label{eq:33}
\snrac^{2}(\Tcoh)\ge\snrth^{2}(\pFA^{*},\pFD^{*})
\end{equation}
holds, where $\snrac^{2}(\Tcoh)$ is the accumulated SNR due to the presence of
signal in the analyzed data. Assuming that the follow-up search will use data
of similar constant noise floor, we can rewrite Eq. (\ref{eq:3}) as
\begin{equation}
 \label{eq:34}
\snrac^{2}(\Tcoh)=\snrcand^{2}\frac{\Nd\Tcoh}{\Ndcand\Tseg}\ ,
\end{equation}
where $\Tseg$ is the length of one segment in the semi-coherent
search using data from $\Ndcand$ number of detectors. With the average
$2\bar{\mF_{c}}$ value of the candidate, we can compute its SNR $\snrcand$
from Eq. (\ref{eq:9}), namely
\begin{equation}
 \label{eq:35}
 \snrcand^{2} = E[2\bar{\mFc}]-4\ .
\end{equation}
Substitution in the equations above yields the accumulated SNR in presence of
signal
\begin{equation}
 \label{eq:36}
 \snrac^{2} = \left(E[2\bar{\mFc}]-4\right)\frac{\Nd\Tcoh}{\Ndcand\Tseg}\ ,
\end{equation}
which gives the required minimal $\Tcoh$.

\subsection{Method 2}

Computation of the $\mF$-statistic on data with no signal, has a certain
expectation value, therefore we ask what is the expected maximal $2\mF$
value $E[2\mFn]$ in $\Ntemp$ trials in Gaussian noise, where
$\mF_{\Ntemp}\equiv\max\left\{\mF\right\}_{i=1}^{\Ntemp}$. The probability to
get $(\Ntemp - 1)$ values of $2\mF$ less than $2\mFn$ follows a binomial
distribution, namely
\begin{eqnarray}
 \label{eq:37-38}
p_{\Ntemp}(2\mFn)&=&
\binom{\Ntemp}{1}\chi^{2}_{4}(2\mF,0)(1-\alpha_1)^{\Ntemp-1}\\
&=&\frac{1}{2}\Ntemp\mFn e^{-\mFn}\left(1-(1+\mFn)e^{-\mFn} \right)^{\Ntemp-1}\
.
\end{eqnarray}
With this we can numerically integrate the expectation value
\begin{equation}
 \label{eq:39}
E[2\mFn] = \int_{0}^{\infty}d(2\mFn)\,2\mFn\, p_{\Ntemp}(2\mFn)\ ,
\end{equation}
and standard deviation
\begin{equation}
 \label{eq:40}
\sigma_{\Ntemp}(2\mFn) = \left(\int_{0}^{\infty}d(2\mFn)\left(2\mFn -
E[2\mFn]\right)^{2}p_{\Ntemp}(2\mFn)\right)^{1/2}\ .
\end{equation}
To safely distinguish a real signal from pure noise, we can require the
following inequality to hold:
\begin{equation}
\label{eq:41}
E[2\mF_{S}]-\sfp\sigma_{S}(2\mF_{S})>E[2\mFn] +
\sfp\sigma_{\Ntemp}(2\mFn)\ ,
\end{equation}
where the expectation value $E[2\mF_{S}]$ of a real signal and its standard
deviation $\sigma_{S}(2\mF_{S})$ are computed using Eqs. (\ref{eq:4}) and
(\ref{eq:5}). As all terms in inequality (\ref{eq:41}) are function of the
observation time, this gives an alternative method to compute the minimal
required integration time. Fine-tuning of Eq. (\ref{eq:41}) is possible through
the safety parameter $\sfp$, which we quantify by using Chebyshev's inequality.
For a random variable $X$, with expected value $E[X]$ and standard deviation
$\sigma$,
\begin{equation}
\label{eq:42}
 P(|X-E[X]|\geq \sfp\sigma)\leq1/\sfp^2\ ,
\end{equation}
which means that at least a fraction 
\begin {equation}
\label{eq:43}
p=1-1/\sfp^2
\end {equation}
 of the data is within $\sfp$ standard deviations on either side of the mean
\cite{sg2000}. Rearranging the above equation yields 
\begin{equation}
\label{eq:44}
 \sfp=1/\sqrt{1-p}\ .
\end{equation}
Having two independent random variables, $2\mF_{S}$ and $2\mFn$, we can label
the fraction of data around each mean as $p_{S}$ and $p_{\Ntemp}$ and introduce
the joint probability $\pjoint=p_{S}p_{\Ntemp}$. We see, that the same joint 
probability can be achieved for different combinations of $p_{S}$ and
$p_{\Ntemp}$. However, a natural choice is $p_{S}=p_{\Ntemp}$, thus 
\begin{equation}
\label{eq:45}
 \sfp=1/\sqrt{1-\sqrt{\pjoint}}.
\end{equation}
We give a set of $\pjoint$ values and related $\sfp$ in Table \ref{Ta_1}.
\begin{table}[h]
\begin{center}
\begin{tabular}{|c|c|c|c|c|}\hline
$\pjoint$ & 0.75 & 0.90 & 0.95 & 0.99\\\hline
$\sfp$ & 2.73 & 4.41 & 6.28 & 14.12\\\hline
\end{tabular}
\end{center}
\caption{Joint probability $\pjoint$ and corresponding required $\sfp$
standard deviations.}
\label{Ta_1}
\end{table}
\\Fixing $\pjoint$ to some value and with this $h$ in inequality (\ref{eq:41}),
we can compute the minimal required coherent observation time $\Tcoh$, such
that (\ref{eq:41}) holds. For this integration time, the joint probability
$\pjoint$ becomes the separation probability $\psep=\pjoint$. This is the
probability, that a candidate due to the presence of a signal is consistent with
the signal hypothesis and a candidate due to the noise is consistent with
the noise hypothesis. Taking into account that $p_{S} = 1 - \pFD$ and
$p_{\Ntemp} = 1 - \pFA$, we find the relation of the separation probability to
the detection probability, namely $\psep = \pdet(1-\pFA)$, or for
negligible false-alarm $\pdet\approx\psep$.

\section{Method comparison}
\label{sec:4}
\subsection{Numerical predictions}
\begin{figure}[htbp]
\centering
\includegraphics[width=0.47\textwidth]{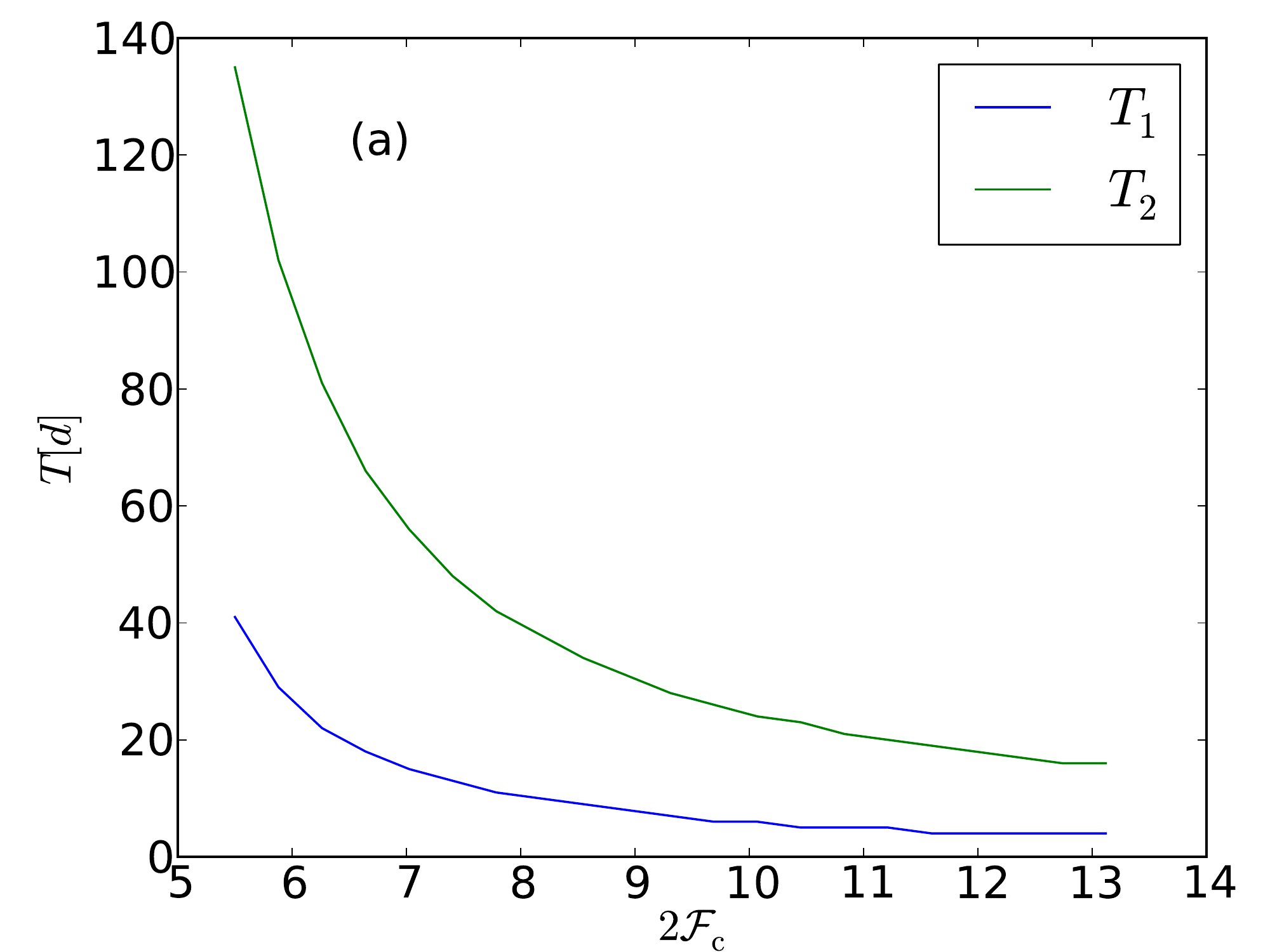}
\includegraphics[width=0.47\textwidth]{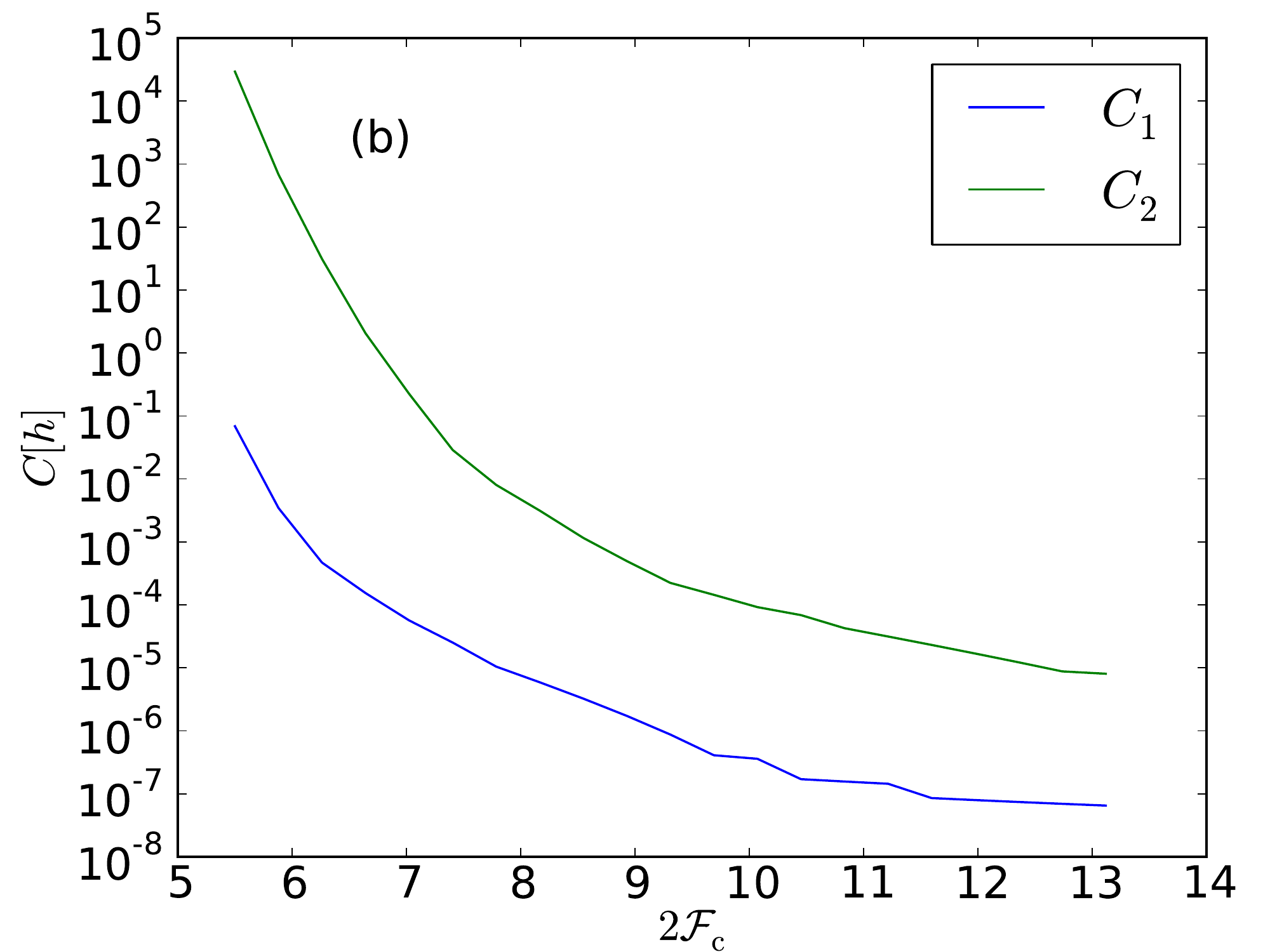}
\includegraphics[width=0.47\textwidth]{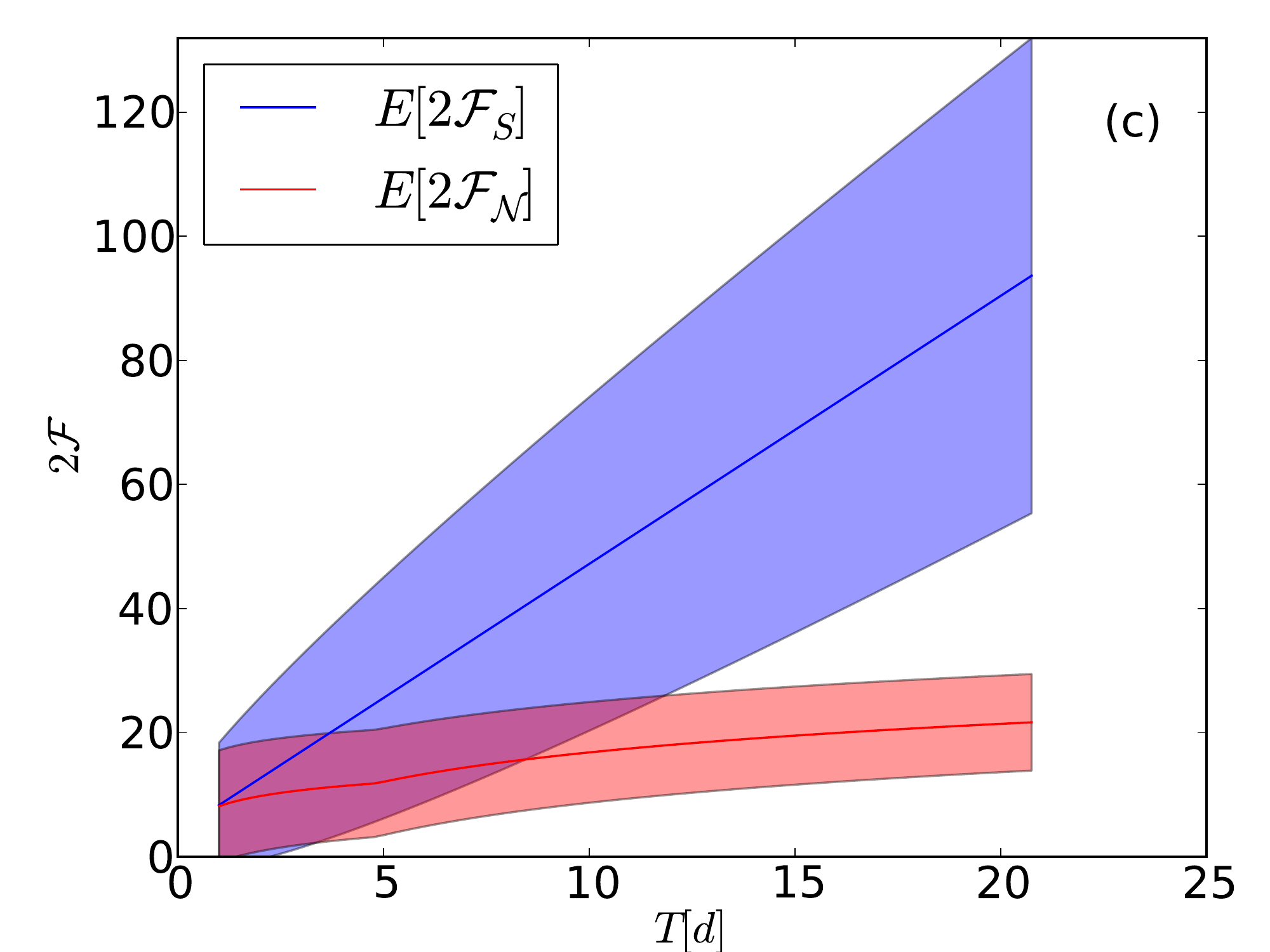}
\caption{Numerical comparison between method 1 and method 2 (quantities labeled
with 1 and 2, respectively). Figure (a) shows the required coherent integration
time as function of the strength of the candidate, (b) shows the computing cost
depending on the strength of the candidate, (c) shows expected value of signal,
noise and related $\sfp=4.41$ standard deviations for detection probability
$\pdet=0.9$ of a candidate with $2\bar{\mF}_{c} = 8.5$\ .}
\label{fig:mecomp}
\end{figure}

In the following we compare the two methods to find the minimal required
integration time described in the previous section in terms of observation
duration and computing cost. We consider a StackSlide search with $\Nseg = 205$
segments of duration $\Tseg=25$ hours, each using data from $\Ndcand=2$
detectors. For a hypothetical candidate with fixed Doppler parameters
$\alpha=1.45\ \radian$, $\delta=0\ \radian$ $f=185\ \Hz$,
$\dot{f}=-1\times10^{-9}\ \Hz/\second$, we pick an average strength in the range
$2\bar{\mF}_{c}\in[5,13]$. Then using Eq. (\ref{eq:15}) with $\colevel=1$ and
the semi-coherent metric we compute the search bands associated with such a
candidate. Having that, for mismatch $m=0.01$ and a hyper-cubic lattice, we can
compute the number of templates using Eq. (\ref{eq:10}) and the fully-coherent
metric. Using method 1, requiring detection probability $\pdet^{*}=0.9$ at
overall false-alarm probability $\pFA^{*}=0.01$ using Eq. (\ref{eq:32}) we
compute $\snrth^{2}(\pFA^{*},\pFD^{*})$ and the minimal required observation
time $\Tcoh_{1}$, which substituted in Eq. (\ref{eq:36}) with $\Nd=\Ndcand$
satisfies Eq. (\ref{eq:33}). For method 2 a separation probability equal
to $\pdet^{*}$ yields safety factor $\sfp = 4.41$, see Table \ref{Ta_1}. We
label the integration time that satisfies Eq. (\ref{eq:41}) as $\Tcoh_{2}$ and
plot both integration times $\Tcoh_{1}(2\bar{\mF}_{c})$ and
$\Tcoh_{2}(2\bar{\mF}_{c})$ in Figure \ref{fig:mecomp} (a) as function of
$2\bar{\mF}_{c}$. With the number of templates for $\Tcoh_{1}$ and $\Tcoh_{2}$
we estimate the computing cost $C_{1}$ and $C_{2}$ using the fundamental
computing cost constant $c_{0}=7\times10^{-8} s$ in Eq. (\ref{eq:26}) and
assuming SFTs of duration $\Tsft=1800\ \second$ in Eq. (\ref{eq:27}).
$C_{1}(2\bar{\mF}_{c})$ and $C_{2}(2\bar{\mF}_{c})$ are plotted in Figure
\ref{fig:mecomp} (b). In Figure \ref{fig:mecomp} (c) we plot how the expectation
value from a real signal grows with increasing $\Tcoh$ compared to loudest
candidate from Gaussian noise. In this plot the candidate strength is fixed to
$2\bar{\mF}_{c} = 8.5$.\\
We see that method 2 yields much longer observation time, at same candidate
strength compared to method 1. Due to the resulting much larger number of
templates, the computing cost, especially for weak candidates, is much higher.
The inferiority of method 2 compared to method 1 in terms of required
integration duration and computing power can be explained by the \textit{ad hoc}
construction of method 2 and the use of Chebyshev's inequality, which is only
a lower bound. In this sense method 2 is a more conservative approach, though
the important information about false-alarm and false-dismissal probability gets
lost in this framework. The computing cost of method 1 looks very promising even
for weak candidates, however we should keep in mind that this is lower limit and
the cost of a search with real data would most likely be much higher. The reason
for this is that gaps in the data are direct penalty for the growth of
$\snrac^{2}$, while $\snrth^{2}$ remains unaffected. Furthermore, for very weak
signals, the required integration duration may violate the assumption of
constant sky resolution, thus we would underestimate the number of templates,
resulting in a higher false-dismissal.

\subsection{Monte Carlo results}
To confirm the numerical predictions of method 1 we perform the following Monte
Carlo studies. We create a set of 205 segments with duration 25 hours of
Gaussian noise and draw a set of pulsar parameters $\alpha\in(0,2\pi)$,
$\delta\in(-\pi/2,\pi/2)$, $\cos\iota\in(-1,1)$, $\psi\in(0,2\pi)$,
$\phi_{0}\in(0,2\pi)$ at fixed frequency of $f=185\ \Hz$ and spindown value in
the range $\dot{f}\in(-f/\tau,0)$, where $\tau=2220\ \years$ is the minimal
spindown age of the source \cite{Brady:1997ji}. We inject a signal with the
above parameters and intrinsic signal amplitude $h_{0}$ high enough to produce a
candidate with expected average strength $E[2\bar{\mF}_{S}]\in[12,13]$. To find
the actual injected value we first do a targeted StackSlide search at the point
of the injection. With this measured injected $2\bar{\mF}_{S}$ value, using Eq.
(\ref{eq:15}) we compute Fisher extents, from which we draw a random
parameter point $\lambda_{c}$ satisfying 
\begin{equation}
 \bar{\Gamma}_{ij}\Delta\lambda^{i}\Delta\lambda^{j}<1\ .
\end{equation}
The point $\lambda_{c}$ is within the $1$-$\sigma$ Fisher ellipsoid of the true
signal location and becomes the candidate to follow up. Following the scheme
for method 1 as described above, we compute the minimal required coherent
observation time targeting detection probability $\pdet^{*}=0.9$ and search for
the signal. After computation of $2\mF_{S}$ using the data with the injected
signal, we compute $2\mF_{\Ntemp}$ with the same grid and integration
duration using the noise only data. We claim ``detection'' whenever the loudest
measured $2\mF_{S}$ value in the data with injected signal is higher than the
loudest measured $2\mF_{\Ntemp}$ of the noise. The result of the Monte Carlo
simulations is as follows: in 897, out of 1000 trials, the measured $2\mF_{S}$
value in the data containing injected signal exceeds the measured
$2\mF_{\Ntemp}$ value of the noise only data. With this the achieved detection
probability $\pdet = 0.897\pm0.023$ is in accordance with the targeted
detection probability $\pdet^{*}$.

\section{Discussion}
\label{sec:5}

We derived two different methods to compute the minimal required coherent
integration time in a fully-coherent $\mF$-statistic search in the \textit{zoom}
stage of follow-up of candidates from a semi-coherent StackSlide search.
By numerical comparison we showed that method 1 is superior to method 2 in
terms of required integration duration and computing cost. We confirmed in a
Monte Carlo study that the predicted coherent integration time is sufficient to
achieve the desired detection probability.
The results of this paper have been derived for Gaussian data without gaps and
two detectors of equal noise floor. Further extension of this work is closely
related to the data selection problem.

\ack
This work significantly benefited from numerous suggestions of Reinhard Prix. I
also thank Holger Pletsch, Karl Wette and Paola Leaci for useful discussions.
Finally I would like to acknowledge the support of Bruce Allen and the IMPRS on
Gravitational Wave Astronomy of the Max-Planck-Society.

\section*{References}
\bibliographystyle{iopart-num}
\bibliography{amaldi_2011_pro}
\nocite{*}

\end{document}